\documentclass[fleqn,10pt]{wlpeerj}

\title{Dissecting the statistical properties of the
        Linear Extrapolation Method 
        of determining protein stability}

\author[1]{Kresten Lindorff-Larsen}
\affil[1]{Structural Biology and NMR Laboratory \& Linderstr{\o}m-Lang Centre for Protein Science, Department of Biology, University of Copenhagen, Copenhagen, Denmark}
\corrauthor[1]{Kresten Lindorff-Larsen}{lindorff@bio.ku.dk}


\begin{abstract}%
The linear extrapolation method to determine protein stability from denaturant-induced unfolding experiments is based on the observation that the free energy of unfolding is often a linear function of the denaturant concentration. The value in the absence of denaturant is then estimated by extrapolation from this linear relationship. Parameters and their confidence intervals are typically estimated by nonlinear least-squares regression. We have compared different methods for calculating confidence intervals and found that a simple method based on linear theory gives accurate results. We have also compared three different parameterizations of the linear extrapolation method and show that the most commonly used form is problematic since the stability and m-value are correlated in the nonlinear least-squares analysis. Parameter correlation can in some cases cause problems in the estimation of confidence-intervals and -regions and should be avoided when possible. Two alternative parameterizations show much less correlation between parameters.\end{abstract}

\begin{document}

\flushbottom
\maketitle
\thispagestyle{empty}
\newpage
\section*{Introduction}
The thermodynamic stability of a protein is an important parameter in
many areas in protein chemistry ranging from biotechnological
applications to basic aspects of protein chemistry.
The stability of a protein
is generally expressed as the change in Gibbs free energy
accompanying the unfolding reaction or equivalently as the
equilibrium constant for this reaction. Therefore it is necessary to
determine the population of both the native and the denatured states
in order to evaluate the stability. Under physiological conditions
most proteins are found almost exclusively in their native state, and
therefore it is in general not possible to estimate directly the
equilibrium constant for the unfolding reaction.

The general solution
to this problem is to perturb the stability of the protein, typically
by either heat or addition of chemical denaturants. Under such
destabilizing conditions both the folded and the denatured states are
populated significantly and thus the ratio of the two concentrations
are within the limits of experimental determination. In practice such
experiments are often carried out by measuring some spectrometric
signal, $\alpha$, such as fluorescence or circular dichroism as a
function of the perturbing variable ($T$: temperature, $P$: pressure
or $\xi$: solvent composition such as by adding of chemical
denaturants or other co-solvents). Under the assumption that the
measured value $\alpha$ is the population-weighted average over all
microstates, and that these microstates have been grouped into two
macrostates, A and B (e.g. native and unfolded), it can be shown \citep{brandts1969conformational,santoro1988unfolding} 
that the functional dependence of $\alpha$ is
given by:

\begin{equation}
  \label{eq:santoro}
  \alpha(T,P,\xi) = \frac%
                        {\alpha_A(T,P,\xi) + \alpha_B(T,P,\xi)exp(-\Delta_rG(T,P,\xi)/RT)}%
                        {1+exp(-\Delta_rG(T,P,\xi)/RT)}
\end{equation}
where $\alpha_A$ and $\alpha_B$ refer to the population-weighted
averages of the spectrometric signal in the A and B macrostates,
respectively, and $\Delta_rG(T,P,\xi)$ is the difference in Gibbs free
energy between the $A$ and $B$ states. All the functions
$\alpha$, $\alpha_A$, $\alpha_B$, and $\Delta_rG$ depend on the
perturbing parameters, $T$, $P$, and $\xi$. When $\Delta_rG$ is large
and positive (native conditions) the measured signal is approximated
well by $\alpha_A$, whereas it is approximately equal to $\alpha_B$
under denaturing conditions (large and negative $\Delta_rG$). Thus,
$\alpha_A$ and $\alpha_B$ are the pre- and post-transition baselines,
respectively.

When the perturbing parameter is the molar denaturant concentration,
$D$, functional dependencies for the three functions $\alpha_A(D)$,
$\alpha_B(D)$, and $\Delta_rG(D)$ are inserted into
Eq.~\ref{eq:santoro} which thereby describes the experimental
measurements of $\alpha$ as a function of $D$ and a number of unknown
parameters. The goal is to determine these parameters from the
experimental data. This is generally done by nonlinear regression to
the parametric version of Eq.~\ref{eq:santoro}. Often the pre- and
post-transition baselines are approximated well by linear functions,
although such choices in general are empirically based.

\begin{figure}[tbp!]
      \centering
      \includegraphics[width=0.4 \columnwidth]{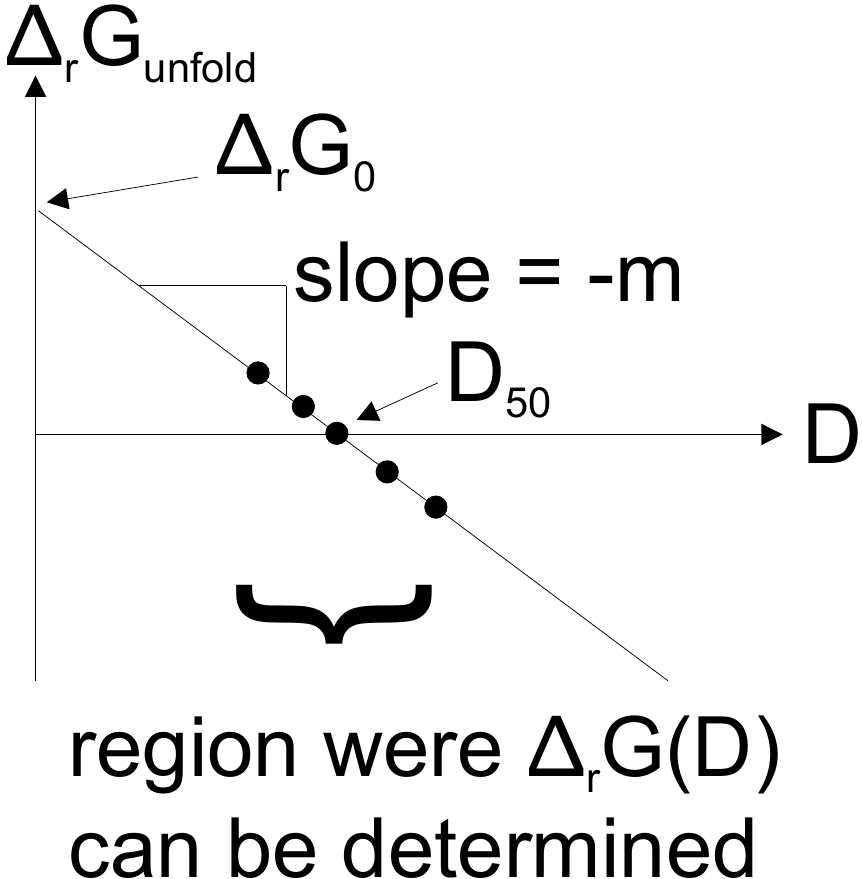}
      \caption{\label{fig:DrG_diagram}
      The linear extrapolation method. The geometric interpretation of the
    three parameters, $\Delta_rG_0$, $m$ and
    $D_{50}$, that are used as parameters to describe the linear relationship
    between unfolding free energy and denaturant concentration is shown. The interval
    of denaturant concentrations which allows a reasonable determination of
    $\Delta_rG$ is sketched.
      }
\end{figure}

Although the thermodynamic theory for denaturant induced unfolding is
highly developed (\cite{schellman1994thermodynamics,schellman2002fifty} and references therein), there exists
no full thermodynamic theory which suggests a suitable and
sufficiently simple parameterization of the dependence of $\Delta_rG$ on $D$.
Experimental evidence, however, points to that the free energy change
associated with the unfolding reaction is often approximated well by
a linear function of the denaturant concentration \citep{tanford1970protein,greene1974urea,pace2000linear}:

\begin{equation}
  \label{eq:LEMEq1}
  \Delta_rG(D) = \Delta_rG_0 - mD
\end{equation}
The minus-sign in Eq.~\ref{eq:LEMEq1} is introduced so that $m$
is a positive quantity (the unfolding free energy decreases as
denaturant is added) and $\Delta_rG_0$ is the value of the free
energy of unfolding in the absent of denaturant (Fig.~\ref{fig:DrG_diagram}).
In the remainder of this paper we assume that this linear relationship is valid
and refer the reader to the literature for examples and discussions
of when the linear model might not hold
\cite{yi1997characterization,schellman2002fifty,moosa2018denaturant,amsdr2019urea}.

This strategy is generally known as the linear extrapolation method
(LEM) \citep{santoro1988unfolding,pace2000linear}. We should note that other functions in addition to a linear
one have been used to parameterize $\Delta_rG(D)$, however these will
not be considered here. The reason that Eq.~\ref{eq:LEMEq1} is an
extrapolation originates from the fact that, as explained above, 
$\Delta_rG(D)$ can only be measured in some interval around $D_{50}$
(the denaturant concentration where half the protein is unfolded and
therefore $\Delta_rG(D_{50})=0$), since only in this interval will
there be sufficient amounts of both native and unfolded protein
present to allow the detection of both. Thus, $\Delta_rG_0$ is effectively an
\emph{extrapolated} quantity.

Since any two of the three parameters $\Delta_rG_0$ (intercept with ordinate), $m$ (the negative of the
slope) and $D_{50}$ (intercept with abscissa) may be used to describe the
linear relationship (Fig.~\ref{fig:DrG_diagram}), an alternative expression is \citep{clarke1993engineered}:

\begin{equation}
  \label{eq:LEMEq2}
  \Delta_rG(D) = -m(D-D_{50})
\end{equation}
This equation can be viewed as a truncated series expansion of
$\Delta_rG(D)$ in $D$ at $D_{50}$. Implicit in this equation is that
$\Delta_rG$ is best estimated in a small interval around $D_{50}$ and
therefore this value is the centre of the expansion. An obvious
relationship between the three parameters in
equations~\ref{eq:LEMEq1} and \ref{eq:LEMEq2} exists:

\begin{equation}
  \label{eq:interdependence}
  \Delta_rG_0 = mD_{50}
\end{equation}
A final version of the LEM using $\Delta_rG_0$ and $D_{50}$ as
parameters is:

\begin{equation}
  \label{eq:LEMEq3}
  \Delta_rG(D) = \Delta_rG_0(1-D/D_{50})
\end{equation}
Although the three different ways of expressing the linear dependence
are of course mathematically equivalent, there may be advantages and drawbacks
of each parameterization. In particular we wanted to compare the estimation
of confidence intervals of the fitted parameters in the three
equations.

Since Eq.~\ref{eq:santoro} is nonlinear in the relevant parameters ($\Delta_rG_0$, $m$ and $D_{50}$) there is no unique
way of estimating confidence intervals. We therefore wanted to
compare different methods for obtaining confidence intervals, and to assess the robustness
of how these are estimated.
Also, since it has been reported that there may be extensive correlation
between $\Delta_rG_0$ and $m$ obtained using Eq.~\ref{eq:LEMEq1} \citep{williams2000monte},
we wanted to explore parameter correlation in all three
parameterizations of the LEM. To this end we have analysed both
synthetic as well as experimental data and found that the parameters in equations
\ref{eq:LEMEq2} and \ref{eq:LEMEq3} are much less
correlated, and these might therefore be preferred. Also, our results show that
a simple method for estimating confidence intervals, and which is used in many software packages for non-linear regression,
is as least as good as more advanced ones.

\section*{Methods}%
\subsection*{Protein unfolding data}%
We chose to use synthetic data  to obtain sets of data with well defined statistical
properties . To mimic the results from
a standard determination of protein stability, data were generated as
follows. First nine sets of noiseless data were generated from
equations \ref{eq:santoro} and \ref{eq:LEMEq2} using every
combination of $m=\{4,6,8\}\ kJ mol^{-1} M^{-1}$ and
$D_{50}=\{3,4,5\}\ M$, thus simulating a range of protein stabilities
between $12$ to $40\ kJ mol^{-1}$. Thirty $D$-values were distributed
between $D=0 M$ and $D=8 M$. Ten of these were distributed
equidistantly in the transition region between $D_{-}$ and $D_{+}$,
chosen to correspond to $K=0.1$ and $K=10$, respectively, where $K$
is the equilibrium constant for the unfolding reaction. The remaining
20 data points were distributed equidistantly between $D=0 M$ and
$D=8 M$. To mimic the situation were both protein unfolding and
refolding experiments are carried out, each data point
was duplicated. The pre- and post-transition baselines were arbitrarily chosen as 
$\alpha_A = 200 + 5D$ and $\alpha_B = 500 + 7D$, 
respectively. Finally, pseudo-random, normally distributed noise with
zero mean and a standard deviation of 10 was added to each of the
nine synthetic data sets. As example of experimental data
we used previously reported measurements for the barley protein LTP1 \citep{lindorff2001surprisingly}.

\subsection*{Numerical methods}%
We used Eq.~\ref{eq:santoro} with linear pre- and post-transition
baselines  in all regression analyses using one of equations
\ref{eq:LEMEq1}, \ref{eq:LEMEq2} or \ref{eq:LEMEq3} as parametric
forms of the LEM, in each case giving a total of 6 parameters  (two for each baseline, and two in the LEM).
Nonlinear least-squares regression was carried out using the Marquardt algorithm \cite{marquardt1963algorithm}.
The Cholesky matrix decomposition was used to
determine step-directions and -lengths. Further, the decomposition
can be used as a test of positive definiteness of the modified
Hessian matrix to ensure that all parameter steps are acceptable
\citep{bard1974nonlinear}. The termination criteria were that (1) the sum-of-squares
function should not decrease by more than $10^{-5}$ and (2) that the
attempted update of
parameter $i$ ($\delta p_i$) should satisfy the equation 
$|\delta p_i| < 10^{-6}(|p_i|+10^{-12})$ for all parameters.

Confidence intervals were estimated using several different methods.
The simplest is the approximate marginal confidence interval
\citep{bates1988nonlinear, seber1989nonlinear} which scales the square root of the diagonal
elements of the variance-covariance matrix using the
\emph{t}-distribution. The variance-covariance matrix is proportional
to the inverse of the curvature matrix for the sum-of-squares
function. Therefore, in effect this method attempts to predict the
behaviour of the sum-of-squares function outside the minimum using information on the
curvature at the position of the best-fit parameters. In the case of
a linear fitting function the sum-of-squares function is second order
and this method is exact. However, with nonlinear equations this is
clearly an approximation. It should be noted that many software
packages for nonlinear regression give the square root of the
diagonal elements of the variance-covariance matrix as estimates of
the uncertainty of the fitted parameters. To obtain confidence
intervals from these values one must manually scale by a $t$-value.

We also used the search method developed by Johnson \citep{johnson1981analysis,johnson1992},
which extends an idea from \cite{box1960fitting}, to evaluate
confidence regions. In contrast to the marginal method described above, this method
does not try to estimate confidence regions from properties at the
minimum of the target function. Instead a search is carried out in
parameter space in carefully chosen directions in order to find
confidence regions. Apart from not only using information at a single
point this method is further strengthened by the fact that intervals
are not necessarily assumed to be symmetric.

Finally we used Monte Carlo procedures to obtain information
regarding the distribution of the fitted parameters \citep{straume19927}. From
the best-fit parameters noiseless data were generated at the same
$D$-values as those in the original data. 500 sets of synthetic data
were generated by adding pseudo-random noise. This was either
generated as pseudo-random normally distributed noise with zero mean
and variance equal to that of the original fit or by a bootstrap
procedure \citep{chernick1999bootstrap}. In the latter, residuals from the fit were
randomly drawn (with replacement) and added to the noiseless data.
Each of the 500 synthetic data sets were analysed by nonlinear
regression to obtain 500 sets of parameters. These were then used to
estimate the variance-covariance matrix. Confidence intervals
were estimated by finding the minimal interval of the 500 parameter
values that contains the appropriate number of points (e.g. 340 of
the 500 in the case of a 68\% confidence interval).

The success of the different methods of estimating confidence
intervals were estimated by the following procedure. 1000 synthetic
dataset were generated from a fixed set of parameters,
$\hat{p}=\{\hat{p}_1,\ldots,\hat{p}_6\}$ by addition of normally
distributed pseudo-random noise with zero mean and standard deviation
10. Confidence intervals for each parameter $p_i$ were calculated for
all 1000 dataset using each of the above mentioned methods. We then
counted the number of times in these fits that the estimated confidence interval for $p_i$
included the original parameter value $\hat{p}_i$. This number was
then compared to the number expected which, e.g. in the case of a
68\% confidence interval, would be 680. To check the robustness of
the methods for estimating confidence intervals we carried out the
same analysis using noise drawn from the Lorentz distribution,
which has longer tails than the normal distribution and
therefore this method simulates experiments with more frequent
outliers. Since neither the mean nor the variance are defined for the
Lorentz distribution we chose the parameters so that the median and
the half-width were the same as in the case of Gaussian noise.
Further we, arbitrarily, chose to cut off noise further away than 200
in order not to get outliers too far away.

In the Monte Carlo analyses and for generation of the synthetic
unfolding dataset we used the Mersenne Twister \citep{matsumoto1998mersenne} as
pseudo-random number generator. In the relevant cases, these were
converted into normally distributed random deviates by the
Box-M\"{u}ller transform \citep{box1958note,knuth2014art}. Lorentzian pseudo-random numbers
were generated directly from the inverse of the distribution
function.

The correlation matrix is a scaled variance-covariance matrix
\citep{johnson2000parameter} and was calculated from:

\begin{equation}
  \label{eq:corrDefinition}
  corr(x_i,x_j) = cov(x_1,x_2)/(\sigma_{x_i}^2 \sigma_{x_j}^2)^{-1/2}
\end{equation}
where $cov(x_1,x_2)$ is the covariance between $x_1$ and $x_2$. The
correlation matrix contains elements with values between $-1$ and
$1$. An absolute value near $1$ indicates a high degree of
correlation between the two parameters during the regression
analysis.

\subsection*{Error propagation}
When a parameter, $p$, is a function of two other parameters,
$p=p(x_1,x_2)$, such as in Eq.~\ref{eq:interdependence} or variants
thereof, standard theory for error-propagation \citep{bevington1993data} shows
that the standard deviation of $p$, $\sigma_p$, can be estimated
using:

\begin{equation}
  \label{eq:ErrorPropagation}
  \sigma_p^2 \approx \sigma_{x_1}^2 (\frac{\partial p}{\partial x_1})^2 +%
                     \sigma_{x_2}^2 (\frac{\partial p}{\partial x_2})^2 +%
                     cov(x_1,x_2) \frac{\partial p}{\partial x_1} \frac{\partial p}{\partial x_2} %
\end{equation}
where it is understood that the partial derivatives should be
estimated at the given values of $x_1$ and $x_2$, and where
$\sigma_{x_1}$ and $\sigma_{x_2}$ are the standard deviations of
$x_1$ and $x_2$ and $cov(x_1,x_2)$ is their covariance.

\section*{Results}
\subsection*{Estimating parameters and confidence intervals using synthetic data}
Since nonlinear least-squares regression is the normal method of
determining the parameters in the LEM as well as their uncertainties,
we wanted to compare the merits of the three parameterizations of the
LEM (Eqs.~\ref{eq:LEMEq1}, \ref{eq:LEMEq2} and \ref{eq:LEMEq3}) in such an analysis.
In particular we wanted to explore the
suitability of each equation to determine confidence intervals for
the two parameters, as well as for the third when estimated from the other two
(e.g. $D_{50}$ when fitting to Eq.~\ref{eq:LEMEq1}).

\begin{figure}[tb]
      \centering
      \includegraphics[width=0.9 \columnwidth]{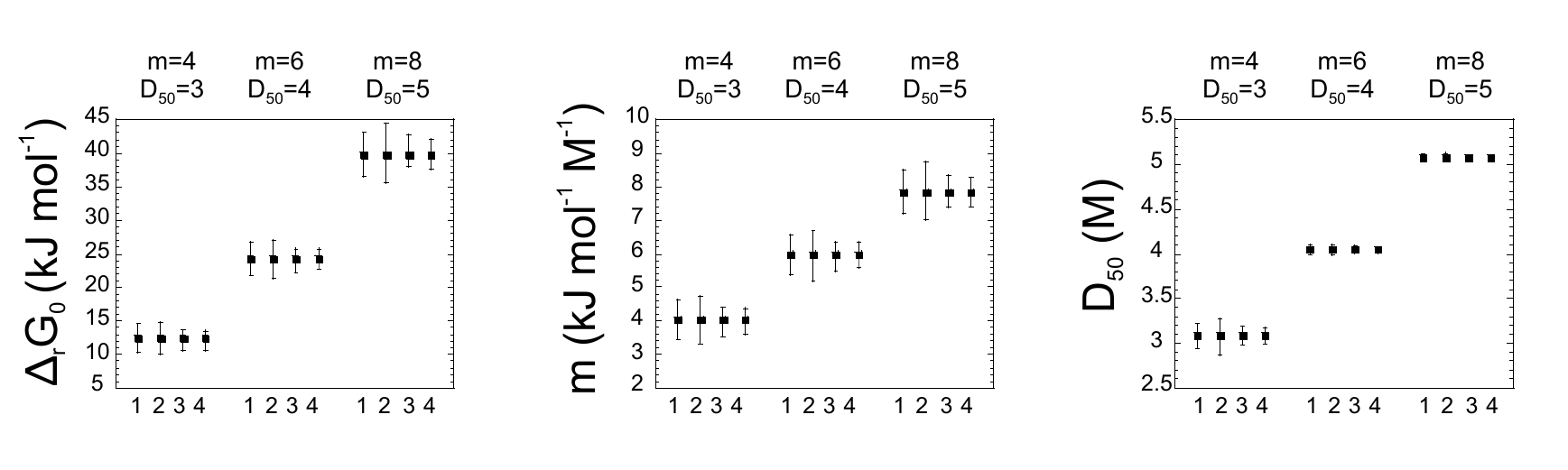}
      \caption{\label{fig:conf_int_methods}
       Comparison of different methods for estimating confidence intervals
    for estimated parameters. Analyses of three of the nine datasets are shown for
    illustrative purposes (indicated by the values of $m$ and $D_{50}$ about each plot). Confidence intervals were calculated as described in the
    methods section using either (1) marginal confidence intervals,
    (2) a search of parameter space for an appropriate variance ratio or Monte Carlo
    methods where either (3) normally distributed noise or a (4) a bootstrap
    procedure was used. The  value on the abscissa indicates which of these methods were used.
     }
\end{figure}

Since all equations are mathematically equivalent
they should all give the same values for the three parameters when
the last (of the three) is calculated using
Eq.~\ref{eq:interdependence}. However, depending on the method
employed, confidence intervals will not necessarily be the same. To
examine this we began by generating synthetic dataset corresponding to
different combinations of protein stability parameters. Thus, we
combined $m$-values of 4, 6 and $8 kJ mol^{-1} M^{-1}$ and $D_{50}$
of 3, 4 and $5 M$ to generate nine dataset corresponding to protein
stabilities ranging from 12 to $40 kJ mol^{-1}$. These dataset were
analysed by nonlinear least-squares regression and 68\% confidence
intervals were estimated for all parameters in the LEM using four
different methods as explained in the methods section. Confidence
intervals from three datasets are shown in
Fig.~\ref{fig:conf_int_methods} for all three parameters of the LEM. Specifically, we fitted to 
Eq.~\ref{eq:LEMEq1} to obtain $\Delta_rG_0$ and $m$, and fitted to Eq.~\ref{eq:LEMEq3}
to obtain $D_{50}$.

While the best-estimated (average) parameters obtained from these methods are consistent,
the different methods for estimating confidence intervals do not give the
same results (Fig.~\ref{fig:conf_int_methods}). Although the differences seem small they were 
consistent within all the analysed data sets. The two different Monte
Carlo methods gave the most narrow confidence intervals, the marginal
method gave intervals that were a bit wider and the parameter space
method gave the widest intervals. The confidence intervals were
almost symmetric for all dataset. Before we analyse which of these approaches give the most 
realistic estimate of the errors, we note that the difference can easily be relevant with
differences in confidence intervals up to two-fold. 
It should be possible to determine small differences in protein stability with accurate measurements,
but interpreting the results require that the confidence intervals are realistic.

We used a Monte Carlo approach to examine which of the methods for estimating confidence intervals
that performed better . In short, we generated 1000
synthetic dataset from a single set of parameters
($\hat{m}=6kJ mol^{-1} M^{-1}$ and $\hat{D}_{50}=4M$). We then estimated  confidence
intervals  for all 1000 datasets using all four methods. For
each method we counted the number of times the parameters $\hat{m}$
and $\hat{D}$ were within the estimated confidence intervals. In the case of,
say, a 68\% confidence interval one would expect about 680 of the
intervals would contain the `true' values. Thus, we compared the
percentages of the confidence level to the number of times $\hat{m}$
and $\hat{D}$ fell within the confidence intervals at both 68\% and
95\% levels. The results (Table~\ref{tab:confIntValidation}) show
that the simple marginal method performed a bit better than the two
Monte Carlo approaches and somewhat better than the parameter space
search method.

To see whether this result would also hold in cases where a robust
confidence estimator method is needed, we repeated the analysis using Lorentzian noise
instead of normally distributed noise. The Lorentz distribution has
much wider tails than the normal distribution and therefore
corresponds to a situation with more frequent outliers. In this case
the least-squares best-fit parameters are not the maximum-likelihood
estimates, however, we were mainly interested in analysing how the
confidence interval estimators performed. The same observation was
observed namely that the marginal method performed at least as good
as any of the others (Table~\ref{tab:confIntValidation}).

\begin{table}[tb] \scriptsize 
 \setlength{\tabcolsep}{2pt}
   \begin{center}
    \caption{\label{tab:confIntValidation}
        Evaluation of confidence intervals by a Monte Carlo method.
    Methods (1) Marginal confidence intervals, (2) parameters space search
    (3) 100 rounds of Monte Carlo analysis, (4) 100 rounds of Monte Carlo bootstrap
        }

  \begin{tabular}{c c c c c c c c c c c c c c c c c c c c c c c c c }

\hline %
Noise       &   \multicolumn{12}{c}{Gaussian}                                                                      &    \multicolumn{12}{c}{Lorentzian}\\ %
Conf. level &    \multicolumn{6}{c}{68.3\%}                 &    \multicolumn{6}{c}{95.0\%}                         &    \multicolumn{6}{c}{68.3\%}                         &    \multicolumn{6}{c}{95.0\%}\\ %
Eq.    & \multicolumn{2}{c}{2} & \multicolumn{2}{c}{3} & \multicolumn{2}{c}{5} & \multicolumn{2}{c}{2} & \multicolumn{2}{c}{3} & \multicolumn{2}{c}{5} & \multicolumn{2}{c}{2} & \multicolumn{2}{c}{3} & \multicolumn{2}{c}{5} & \multicolumn{2}{c}{2} & \multicolumn{2}{c}{3} & \multicolumn{2}{c}{5}\\ %
Parameter   &   m      & $\Delta_rG_0$   & m      & $D_{50}$    & $D_{50}$    & $\Delta_rG_0$   & m     & $\Delta_rG_0$    & m      & $D_{50}$    & $D_{50}$    & $\Delta_rG_0$   &   m    & $\Delta_rG_0$   & m      & $D_{50}$    & $D_{50}$    & $\Delta_rG_0$   & m      & $\Delta_rG_0$   & m      & $D_{50}$    & $D_{50}$    & $\Delta_rG_0$\\ %
Method\\ %
1           &   0.65  & 0.66  & 0.66  & 0.70  & 0.70  & 0.66  & 0.94  & 0.94  & 0.94  & 0.96  & 0.96  & 0.94  & 0.70  & 0.70  & 0.80  & 0.70  & 0.70  & 0.70  & 0.95  & 0.95  & 0.95  & 0.95  & 0.95  & 0.95\\ %
2           &   0.90  & 0.87  & 0.90  & 0.89  & 0.90  & 0.87  & 0.97  & 0.96  & 0.98  & 0.97  & 0.97  & 0.96  & 0.88  & 0.84  & 0.88  & 0.87  & 0.86  & 0.86  & 0.94  & 0.92  & 0.95  & 0.96  & 0.95  & 0.93\\ %
3          &   0.64  & 0.68  & 0.64  & 0.65  & 0.64  & 0.68  & 0.91  & 0.90  & 0.91  & 0.93  & 0.93  & 0.90  & 0.73  & 0.74  & 0.73  & 0.63  & 0.63  & 0.74  & 0.91  & 0.92  & 0.91  & 0.90  & 0.90  & 0.92\\ %
4          &   0.60  & 0.59  & 0.60  & 0.62  & 0.62  & 0.59  & 0.89  & 0.89  & 0.89  & 0.91  & 0.91  & 0.89  & 0.61  & 0.61  & 0.61  & 0.59  & 0.59  & 0.61  & 0.92  & 0.93  & 0.92  & 0.91  & 0.91  & 0.93\\ %
    \hline
  \end{tabular}
  \end{center}
\end{table}

\subsection*{Parameter correlation leads to problems in error propagation}%
Another situation arises when confidence intervals are estimated for
parameters derived from those in the LEM. A typical example is the
calculation of $\Delta_rG_0$ from published values for $m$ and
$D_{50}$ using Eq.~\ref{eq:interdependence}. The standard deviation
of the derived parameter can in general be estimated using the
error-propagation equation (Eq.~\ref{eq:ErrorPropagation}) provided
that sufficient information is available. Since parameters in the
literature are generally presented as $x_i \pm \sigma_{x_i}$ only
information regarding the first two terms in
Eq.~\ref{eq:ErrorPropagation} is available.

\begin{figure}[tb]
      \centering
      \includegraphics[width=0.4 \columnwidth]{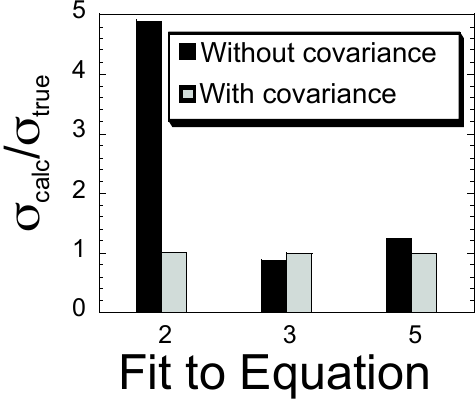}
      \caption{\label{fig:ErrorPropagation}
      Parameter correlation makes it difficult to propagate errors.
    Standard deviations were calculated from the error propagation equation
    (Eq.~\ref{eq:ErrorPropagation})
    using either only the first two terms (black bars) or all three terms (grey bars), i.e. including the covariance.
    The height of the bar indicates the ratio between the calculated standard deviation and that
    obtained directly from nonlinear least squares regression. The results shown
    here were generated from the $m=4\ kJ mol^{-1} M^{-1}$ and $D_{50}=3M$ dataset,
    but all gave similar results.
    Calculations were carried out using
    all three versions of the LEM, so that for Eq.~\ref{eq:LEMEq1} the derived
    parameter is $D_{50}$, for Eq.~\ref{eq:LEMEq2} it is $\Delta_rG_0$ and for Eq.~\ref{eq:LEMEq3}
    $m$ is the derived parameter. It is seen that for Eq.~\ref{eq:LEMEq1}, it is essential to know the
    covariance in order to calculate the standard deviation of the last parameter.
    The other two versions do not suffer from this problem.
      }
\end{figure}

We therefore examined
whether standard deviations calculated using only the first two terms
were reasonable compared to using the full expression (Fig.~\ref{fig:ErrorPropagation}).
For example, we first fitted to Eq.~\ref{eq:LEMEq1} to obtain best fit values of $\Delta_rG_0$
and $m$ as well as the error estimates and covariance between these parameters. We then used 
Eq.~\ref{eq:ErrorPropagation} with either just the two first terms (variance alone) or all three (including covariance)
to estimate the error of $D_{50}$. It is clear that the last term of the error-propagation equation
cannot be ignored meaning that the covariance between the fitted parameters is large.
Thus, while one can estimate errors on $\Delta_rG_0$ from those on $D_{50}$ and $m$ \citep{clarke1993engineered},
one cannot obtain accurate errors in $D_{50}$ from $\Delta_rG_0$  and $m$.

The above observations suggests that this particular parameterization of the LEM
suffers from parameter correlation during the nonlinear least-squares
regression. Parameter correlation is a mathematical feature
indicating that the determination of one parameter by the nonlinear
least-squares procedure is not independent of the determination of
another. The reason is that an increase in the sum-of-squares target
function when one parameter changes is partially compensated by a
change in the correlated parameter. It is important to note that
correlation of this kind has nothing to do with a physical
correlation. To quantify the degree of correlation between parameters
in the three different version of the LEM we calculated correlation
coefficients (Eq.~\ref{eq:corrDefinition}) for all nine dataset
and using all three parameterizations. The average correlation
coefficient between $\Delta_rG_0$ and $m$ in Eq.~\ref{eq:LEMEq1} is
0.99 with the lowest observed value being 0.98. This indicates a very
high degree of correlation. The other two versions of the LEM do not
suffer from the same problem. The average correlation between $m$ and
$D_{50}$ in Eq.~\ref{eq:LEMEq2} is 0.1 with values ranging from -0.5
to 0.6. In Eq.~\ref{eq:LEMEq3} the average correlation coefficient
between $\Delta_rG_0$ and $D_{50}$ is 0.2 (-0.4 to 0.7). For these reasons,
error propagation using just the variance terms are almost the same as also including the
covariance (Fig.~\ref{fig:ErrorPropagation}).

\begin{figure}[tb]
      \centering
      \includegraphics[width=0.9 \columnwidth]{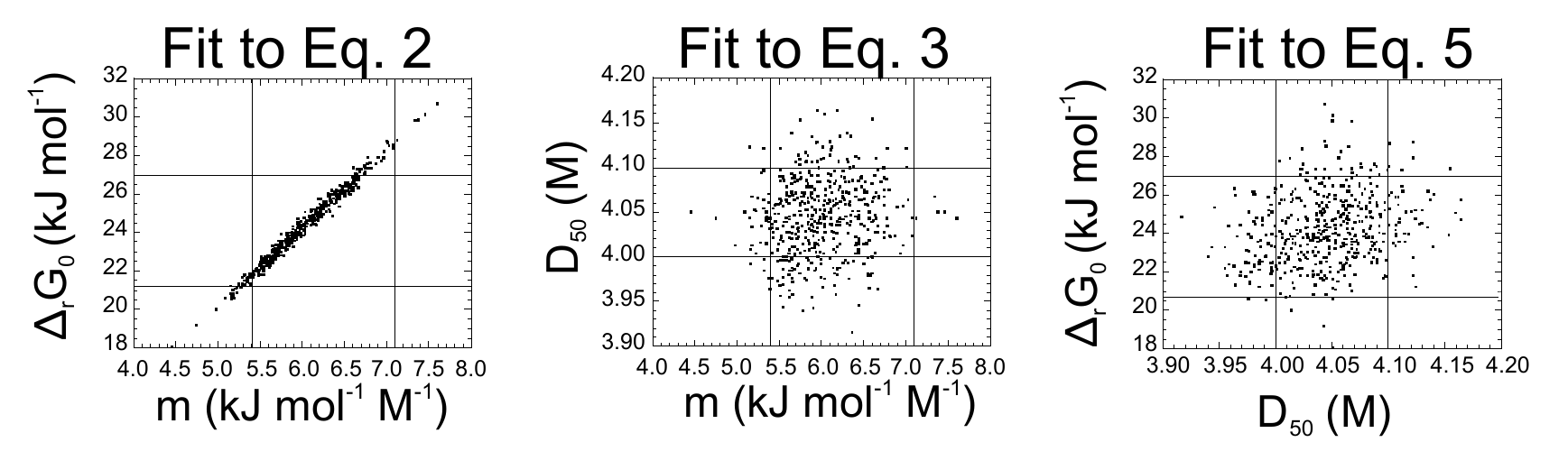}
      \caption{\label{fig:MonteCarlo}
       Parameter distributions
    estimated by a Monte Carlo analysis. We carried out 500 rounds of Monte Carlo simulation
    by generating noiseless data from the best-fit parameters
    and subsequently adding normally distributed noise. Each dataset
    was analysed by nonlinear least-squares and parameters estimated. This was
    repeated for all three parameterizations of the LEM and the figure shows the result
    by plotting each pair of parameters against each other. Note how the parameters
    obtained using Eq.~\ref{eq:LEMEq1} are highly correlated. The horizontal and vertical
    lines indicate 68\% confidence intervals for each parameter as estimated by the marginal method.
      }
\end{figure}

The issue of parameter correlation was further studied by a Monte Carlo
analysis (Fig.~\ref{fig:MonteCarlo}). We generated 500 sets of synthetic data  and each was
analysed by nonlinear least-squares using one of equations
\ref{eq:LEMEq1}, \ref{eq:LEMEq2} or \ref{eq:LEMEq3}. For each parameterisation of the LEM, we 
plot  the value of the two parameters that resulted from these fits (Fig.~\ref{fig:MonteCarlo}). 
The results show very clearly that in the case of
Eq.~\ref{eq:LEMEq1}, the value obtained of one parameter is highly
dependent of that of the other. Also, it is evident that equations
\ref{eq:LEMEq2} and \ref{eq:LEMEq3} do not suffer as badly from this
problem.

The Monte Carlo analysis also highlights a specific issue with the notion
of confidence intervals when parameters are highly correlated.
The lines in the plots
indicate 68\% confidence intervals of each parameter as obtained from
the marginal method (Fig.~\ref{fig:MonteCarlo}).
In the case of Eqs.~\ref{eq:LEMEq2} and \ref{eq:LEMEq3}, the area
covered jointly by the individual confidence intervals correspond quite accurately
to the region found by the Monte Carlo analysis.
On the other hand, for Eq.~\ref{eq:LEMEq1} it is clearly
evidence that the knowledge of the individual confidence intervals of
$\Delta_rG_0$ and $m$ is not sufficient to generate the joint
confidence \emph{region} for the two parameters. Rather, the
confidence region spanned by the two individual confidence intervals
(the inner square) grossly overestimates the area of the confidence
region.This is one of the major reasons why correlated parameters should be avoided
when possible \citep{williams2000monte, johnson2000parameter}. The observation that the
individual confidence intervals are insufficient to determine the
joint confidence region is exactly the reason for why covariances are
necessary to calculate the standard deviation of $D_{50}$ using the
error-propagation equation (Fig.~\ref{fig:ErrorPropagation}).

\begin{figure}[tb]
      \centering
      \includegraphics[width=0.9 \columnwidth]{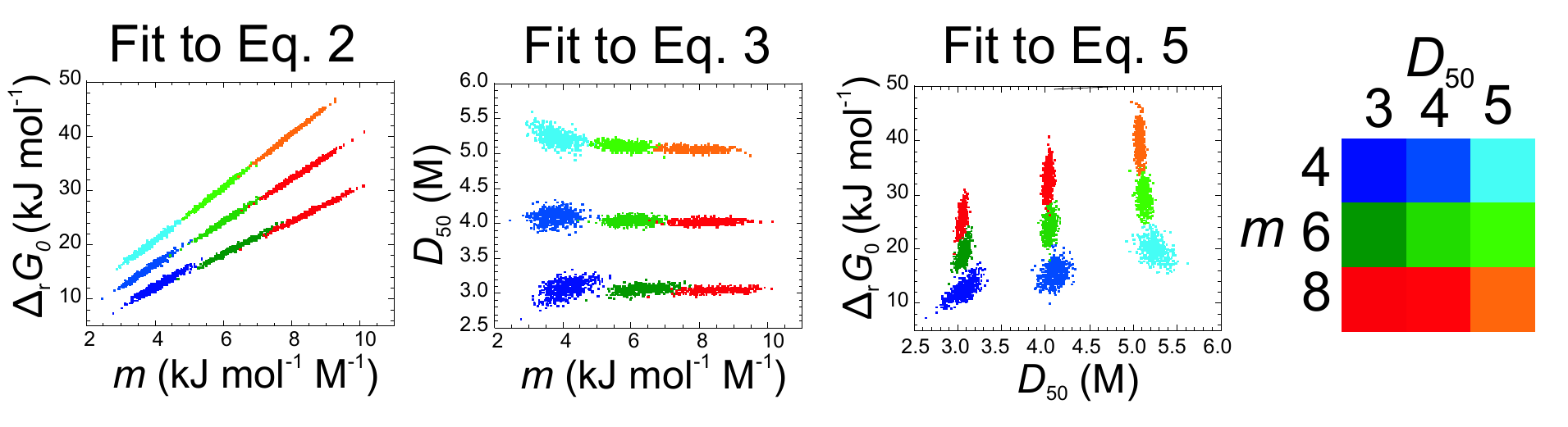}
      \caption{\label{fig:ScatterPlot}
        Comparison of protein stability parameters using scatter plots.
    We carried out a Monte Carlo analysis for all nine synthetic datasets. Each pair of parameters are plotted
    against each other. The colour coding at the right shows the $m$ and $D_{50}$
    values used to generate the original dataset.}
\end{figure}

Plots like those in Fig.~\ref{fig:MonteCarlo} may also be used to
compare denaturation data either under different experimental
conditions or for different mutants \citep{williams2000monte}. An example is shown
in Fig.~\ref{fig:ScatterPlot} where stability parameters for all nine
synthetic dataset are plotted as scatter plots from a Monte Carlo analysis. 
The plots are two-dimensional projections of the three-dimensional
parameter space spanned by $\Delta_rG_0$, $m$ and $D_{50}$ and give
an alternative way of comparing the unfolding behaviour of e.g. a set
of mutants. This type of plot gives an simple way of comparing both the
value of the parameters and their confidence intervals, and may reveal
differences that might otherwise not have been noticed.

Another way of illustrating the correlation between the parameters
uses so-called profile traces \citep{bates1988nonlinear}. These are obtained by fixing one of
the parameters at some given value and estimating the others by
nonlinear least-squares analysis. In the case of non-correlated
parameters, the value of one parameter should not depend on that of
the other. If this procedure is repeated at a number of values for
the fixed parameter and the two parameters are then plotted against
each other, one would then expect an approximately straight line,
parallel to the axis of the fixed parameter. If this is repeated with
the other parameter fixed and subsequently all data are plotted, the
result would be two orthogonal lines each parallel to a parameter
axis. In contrast, in the case of correlated parameters, the value of one parameter
would be dependent of that of the other, and therefore the two curves
would neither be independent nor parallel to either axis. In fact,
the slope of the line is determined by the correlation coefficient
for the two parameters and equal slopes for the two lines are 
obtained in the case of $corr(p_1,p_2)=\pm 1$.

\begin{figure}[tb]
      \centering
      \includegraphics[width=0.9 \columnwidth]{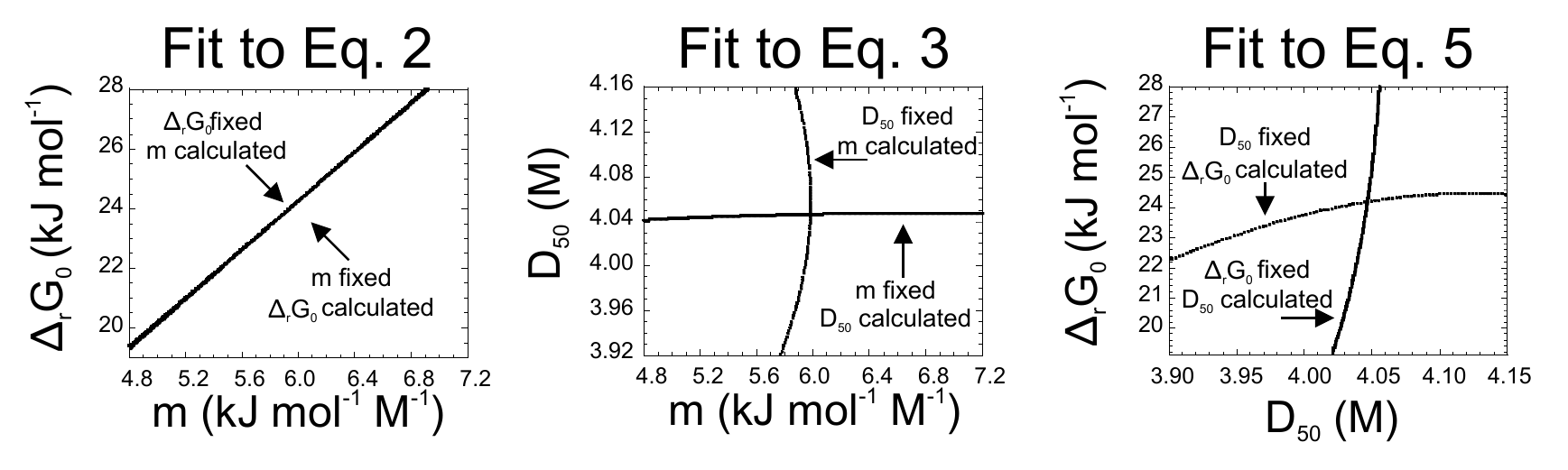}
      \caption{\label{fig:FixedParameters}
       Profile traces \citep{bates1988nonlinear} of parameters in the linear extrapolation methods.
    The denaturation data were analysed by fixing one of
    the parameters in the LEM and estimating the other by nonlinear least-squares.
    This was repeated at several fixed values for both parameters.
    The results shown here are from one of the nine synthetic data set. Similar results
    were obtained from the other eight.
    Two independent lines, as observed for equations
    \ref{eq:LEMEq2} and \ref{eq:LEMEq3}, show that the parameters in these equations
    are not very correlated.
    The curvature of the lines is related to the nonlinearity of the problem. In contrast,
    for Eq.~\ref{eq:LEMEq1} the two lines superimpose. This is a consequence of the correlation
    between $\Delta_rG_0$ and the $m$-value.
      }
\end{figure}

We determined profile traces  for the three versions of
the LEM using one of the synthetic dataset (Fig.~\ref{fig:FixedParameters}).
Again, parameters obtained using Eq.~\ref{eq:LEMEq1} are seen to be
correlated since the change of one parameter is accompanied by a
change in the other. Therefore the two curves coincide. This is not
the case for the other two equations. It should be noted that only in
the case of a strictly linear model will the lines be straight and in
fact it can be shown that the curvature of the lines is an estimate for the
nonlinearity of the model \citep{bates1988nonlinear}.

In addition to visualising the parameter coupling in Eq.~\ref{eq:LEMEq1}, the profile traces are
also relevant to analyses of experimental data in other way. In particular, because $m$ and $\Delta_rG_0$ are correlated,
a common approach for estimating changes in stabilities between e.g. wild type and a mutant ($\Delta\Delta_rG_0$) uses
a fixed (shared) $m$ value for both proteins and estimates
$\Delta\Delta_rG_0 \approx \langle m \rangle (D_{50}^\textrm{mut}-D_{50}^\textrm{wt})$ \citep{kellis1989energetics}.
The profile traces validates this approach by showing that the estimated value of $D_{50}$
is rather insensitive to any noise that might be present when fitting the $m$-value, so that as long as the assumption
that the $m$ value is the same for wild type and mutant is valid, then this approach should indeed be highly accurate.

\subsection*{Analysing experimental data}%
The analyses described above used synthetic data to demonstrate substantial parameter correlation between   
$\Delta_rG_0$ and the $m$-value when these are obtained from fits to 
Eq.~\ref{eq:LEMEq1} . To examine the extent to which these issues also pertain to
experimental data we repeated some of these analyses using previously measured denaturation data
for the protein LTP1 at three different values of pH (pH 3.2, 5.1 and 8.5) \citep{lindorff2001surprisingly}. As with the
synthetic data we find substantial correlations between the fitted parameters when Eq.~\ref{eq:LEMEq1}  is used,
but also sizeable correlations between $m$ and $D_{50}$ in Eq.~\ref{eq:LEMEq2}  (Table~\ref{tab:corrCoef}).

\begin{table}[tb]
   \begin{center}
    \caption{Correlation coefficient between parameters in the
            three versions of the linear extrapolation method.
            \label{tab:corrCoef}
            }
  \begin{tabular}{c c c c c}
    Eq.     &   pH 3.2      &       pH 5.1      &       pH 8.5      & Parameters\\ %
    \hline
   \ref{eq:LEMEq1}        &   0.97        &       0.99        &       1.0         & $\Delta_rG_0 , m$\\ %
   \ref{eq:LEMEq2}        &   -0.55       &       -0.52        &       -0.20      & $m , D_{50}$\\ %
   \ref{eq:LEMEq3}        &   -0.34        &       -0.40        &       -0.11     & $\Delta_rG_0 , D_{50}$\\ %
    \hline
  \end{tabular}
  \end{center}
\end{table}

\begin{figure}[tb]
      \centering
      \includegraphics[width=0.9 \columnwidth]{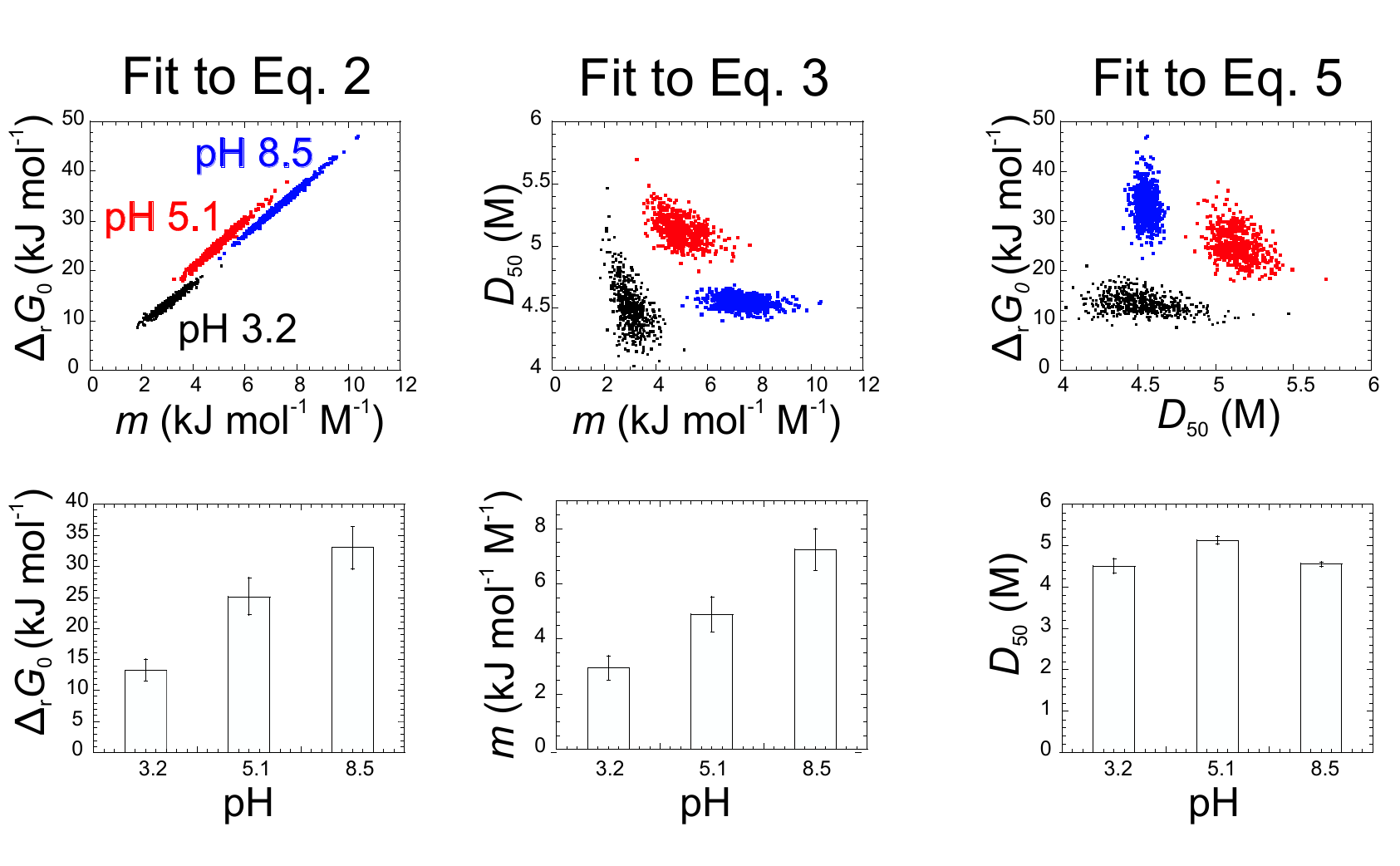}
      \caption{Comparison of protein stability parameters using scatter plots.
                Top row: Data are from LTP1 at pH 3.2 (black), 5.1 (red) and 8.5 (blue).
                Each pair of parameters in the LEM are compared. Parameters were generated by a
                Monte Carlo procedure as in Fig.~\ref{fig:ScatterPlot}. In the bottom row
                data are presented as bar diagrams for each parameter. The standard deviation
                of each parameter is shown as an error bar.
                \label{fig:LTP1ScatterPlot}
                }
\end{figure}

This correlation is also evident when repeating the analysis from the synthetic data (Fig.~\ref{fig:ScatterPlot})
using the data from LTP1 (Fig.~\ref{fig:LTP1ScatterPlot}). In line with the calculated correlation coefficients (Table~\ref{tab:corrCoef})
the plots show varying levels of correlations with strong correlations when fitting to Eq.~\ref{eq:LEMEq1}, intermediate correlations
from Eq.~\ref{eq:LEMEq2} and weak correlations from Eq.~\ref{eq:LEMEq3}. In addition we note that analysing two-dimensional projections
of the fitted parameters (Fig.~\ref{fig:LTP1ScatterPlot}, top) provides a clearer separation of the fitted parameters
at the different values of pH, compared just just examining the individual fitted values (Fig.~\ref{fig:LTP1ScatterPlot}, bottom).
Further, the higher value of $D_{50}$ at pH~5.1 (5.1~M) compared to the values at  pH~3.2 and 8.5 (4.5~M and 4.6~M, respectively) manifests
itself as a slight upward shift in the distribution of $\Delta_rG_0$ and $m$ when fitting to Eq.~\ref{eq:LEMEq1}.

\begin{figure}[tb]
      \centering
      \includegraphics[width=0.9 \columnwidth]{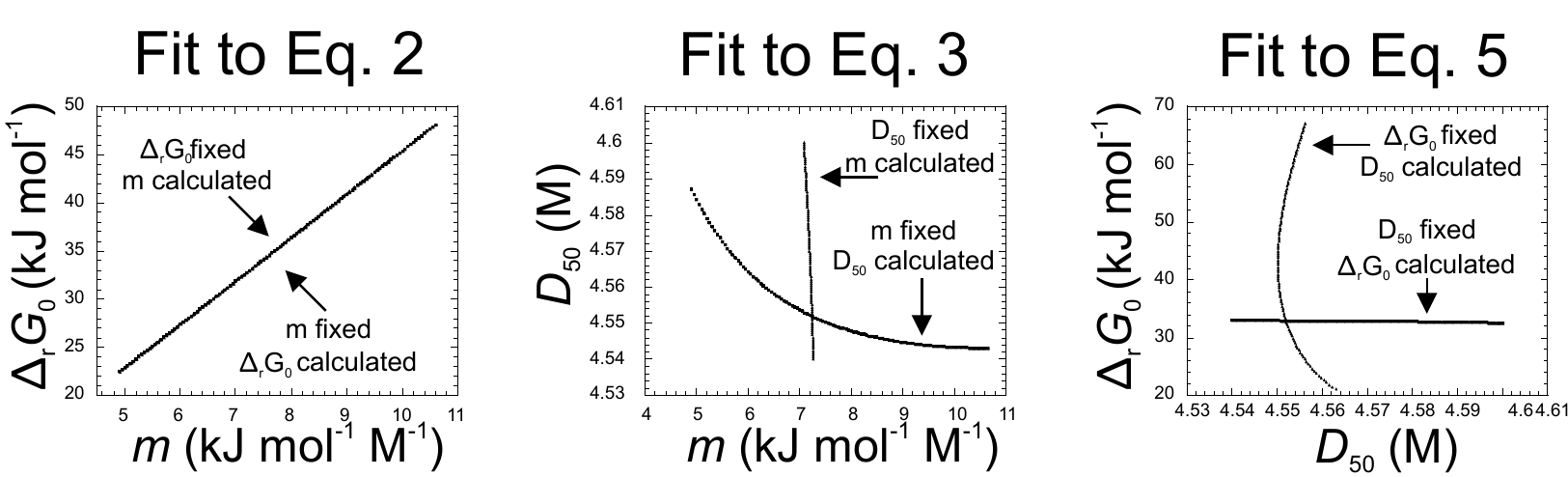}
      \caption{
      Profile traces \citep{bates1988nonlinear} of parameters fitted to
      the LTP1 denaturation data at pH 8.5. As for the synthetic data, two independent lines, as observed for equations
                \ref{eq:LEMEq2} and \ref{eq:LEMEq3}, show that these parameters are not strongly correlated.
                The curvature of the lines is related to the non-linearity of the problem. In contrast,
                for Eq.~\ref{eq:LEMEq1} the two lines superimpose, which is a consequence of the correlation
                between $\Delta_rG_0$ and the $m$-value.
                \label{fig:FixedParametersLTP1}
                }
\end{figure}

We also generated profile traces \citep{bates1988nonlinear} using the LTP1 data at pH 8.5 as an example (Fig.~\ref{fig:FixedParametersLTP1}). Again, 
the strong correlation is evident when Eq.~\ref{eq:LEMEq1} is used so that when e.g. the $m$-value is fixed (and scanned in steps) the 
$\Delta_rG_0$ changes in full concert. In contrast, when Eq.~\ref{eq:LEMEq2} is used and the $m$-value is again scanned, the resulting $D_{50}$
value depends much less on the choice of $m$.

\section*{Discussion}
A major goal in quantative analysis of experimental data is to obtain
estimates of relevant parameters as well as estimates of what level
of confidence in those parameters one should have. In the case of
parameter estimation by nonlinear regression, many software packages 
estimate errors and confidence intervals use the variance-covariance matrix.
Because that method is only strictly accurate for linear equations,
it is not obvious whether that or some of the other methods for estimating fitting errors
would be the best in the case of protein stability
measurements. We therefore compared several different methods for
obtaining confidence intervals using synthetic data with well defined
statistical properties. Our results show that although the confidence
interval estimates are similar, they may differ by a factor of two when obtained by
different methods. We therefore evaluated the performance of four
different methods using a large set of synthetic data. Our results
show that the simplest method performs the best (Table~\ref{tab:confIntValidation}). Further, this is
even the case when non-normal experimental noise is simulated from a
distribution function with wide tails. We therefore suggest that for
most standard analyses, confidence intervals estimated by this method
will suffice --- a fortunate conclusion since this is also the method implemented
in most software packages.
In some cases it may, however, be appropriate to
complement this estimate with a Monte Carlo estimate, in particular
in the case of very noisy data.

We also examined the known problem of parameter correlation when
the fitting protein stability data using Eq.~\ref{eq:LEMEq1} \citep{williams2000monte}.
We note that similar issues have also been noted in the analysis of data from protein folding kinetics
\citep{ruczinski2006methods}.
Our results show that  the minimum of the sum-of-squares
function when using Eq.~\ref{eq:LEMEq1} is long and narrow and is not
aligned with the parameter axes.
Several problems exist when fitting equations with correlated parameters. These include the
problems of constructing joint confidence intervals and in the
application of statistical tests. This is unfortunate since
Eq.~\ref{eq:LEMEq1} uses $\Delta_rG_0$ and the $m$-value as
parameters. These two parameters are those of the three that are
generally of greatest interest because (i) the extrapolated value
$\Delta_rG_0$ is generally assumed to be a good estimate of the
unfolding free energy in the absence of denaturant and (ii) the
$m$-value is known to correlate with changes in accessible surface
area during unfolding \citep{myers1995denaturant,geierhaas2007bppred}.

We, however, also show that two alternative parameterizations of the LEM,
Eq.~\ref{eq:LEMEq2} and \ref{eq:LEMEq3}, suffer much less from parameter
correlation, likely because they include $D_{50}$ which is the best determined of the parameters.
Whereas a small change in $\Delta_rG_0$ may be compensated by one in
$m$ in Eq.~\ref{eq:LEMEq1}, this is not possible for the other
equations since $D_{50}$ is determined much better by the
experimental data. Similarly, if $D_{50}=\Delta_rG_0 / m$ is the parameter
best determined by the experiments, it is clear that $\Delta_rG_0$ and $m$ will be positively
correlated. We show that due to parameter correlation, it is not possible to estimate
confidence intervals for $D_{50}$ from the confidence intervals of $\Delta_rG_0$ and $m$,
and suggest that it is safest also to specify $D_{50}$ explicitly.

Finally, our analyses support the use of a variant of Eq.~\ref{eq:LEMEq2} to estimate
changes in stability between two protein variants. Our profile trace analysis
show that $D_{50}$ can be robustly determined even if the $m$ value is not
accurate, supporting the common practice of combining an estimate of the mean $m$ value and changes
in $D_{50}$ to estimate changes in protein stability.
\section*{Acknowledgments}
K.L.-L. acknowledges Professors Jane Clarke and Christopher M. Dobson for discussions and input to this work.
K.L.-L. was supported by the Danish Research Agency and is supported by the
Novo Nordisk Foundation.
\bibliography{references}

\end{document}